\documentclass[a4paper]{jpconf}
\usepackage{graphicx}
\begin{document}
\title{Fusion barrier distribution and superheavy elements}

\author{K. Hagino$^{1,2}$ and T. Tanaka$^{3,4}$}

\address{$^1$ 
Department of Physics, Tohoku University, Sendai 980-8578, Japan}
\address{$^2$  
Research Center for Electron Photon Science, Tohoku University,
1-2-1 Mikamine, Sendai 982-0826, Japan}
\address{$^3$ RIKEN Nishina Center for Accelerator-Based Science, Saitama 351-0198, Japan}
\address{$^4$ Department of Physics, Kyushu University, Fukuoka 819-0395, Japan}


\begin{abstract}
The barrier distribution, originated from channel coupling effects in 
heavy-ion fusion reactions, 
has been extracted experimentally 
for many systems using either a fusion excitation function or an excitation 
function of large-angle quasi-elastic scattering. 
In this article, we discuss an application of the latter method to the $^{48}$Ca+$^{248}$Cm 
system, which is relevant to 
hot fusion reactions to synthesize superheavy elements. 
To this end, we carry out 
coupled-channels calculations for this system, 
taking into account the deformation of the 
target nucleus, and 
discuss the role of deformation in a formation of 
evaporation residues. 
\end{abstract}

\section{Introduction}

Fusion reactions play an important role in several phenomena in physics. 
Those include the energy production in stars, nucleosyntheses, and 
formations of superheavy elements. 
Yet, from the theoretical point of view, fusion reactions, as well as fission, are typical examples of large amplitude collective 
motions of quantum many-body systems, 
and their microscopic understanding has still been 
far from complete. 

In order to understand the dynamics of nuclear fusion, the Coulomb barrier between two 
nuclei plays an important role.  This is a 
potential barrier formed as a result of cancellation between the long-ranged
repulsive Coulomb interaction and a short ranged attractive 
nuclear interaction. 
The height of the Coulomb barrier defines the energy scale of a system, and  
here in this article, we shall mainly consider the energy region 
around the Coulomb barrier. 

It has by now been well known that heavy-ion fusion cross sections are largely enhanced 
at energies below the Coulomb barrier, as compared to a prediction of 
a simple potential model. 
It has been understood that channel coupling effects, that is, the 
couplings of the relative motion between two nuclei to several collective 
excitations of the colliding nuclei, as well as particle transfer channels, 
play an essential role in enhancing subbarrier fusion cross sections 
 \cite{BT98,DHRS98,HT12,Back14,MS17}. 
The coupled-channels approach has been developed in order to take into 
account such channel coupling effects \cite{HRK99}. 

In the eigen-channel representation of the coupled-channels method, 
fusion cross sections in the presence of channel couplings can be 
represented as a weighted sum of fusion cross sections for 
each eigen-channel \cite{DLW83}. 
In this picture, a single Coulomb barrier is replaced by a distribution 
of multitude of barriers due to the channel coupling effects. A way to 
extract the barrier distribution directly from experimental fusion cross 
sections has been proposed by 
Rowley, Satchler, and Stelson \cite{RSS91}. In this method, the barrier distribution 
is extracted by taking the second energy 
derivative of the product of fusion cross sections, $\sigma_{\rm fus}$, 
and incident energies in the center of mass frame, $E$, that is, $d^2(E\sigma_{\rm fus})/dE^2$. 
The fusion barrier distribution has been extracted with this method for 
many systems \cite{DHRS98,LDH95}, which in general 
show that the shape of barrier 
distribution is sensitive to details of channel couplings, thus providing 
a good tool to understand the underlying dynamics of subbarrier fusion 
reactions. 

It has been shown that a similar barrier distribution can be obtained 
also by using quasi-elastic scattering at backward angles \cite{TLD95,HR04}. 
This is because quasi-elastic scattering corresponds to a reflection of 
flux at the barrier and thus it is complementary to fusion, which corresponds 
to a transmission of flux through the barrier. 
Very recently, the quasi-elastic barrier distribution was extracted 
for the $^{48}$Ca+$^{248}$Cm system \cite{Tanaka18}, the system 
which had been used to synthesize superheavy 
elements $Z=116$ (Lv) \cite{Oganessian04,Hofmann12}. 
Using the GARIS ion separator at RIKEN, which has been used to synthesize 
the $Z=113$ element (Nihonium) \cite{Morita12}, quasi-elastic cross sections
have been successfully measured, which are almost 
free from contamination of deep-inelastic cross sections. 
The barrier distribution extracted from such data is thus much better 
defined as compared to the previous attempts \cite{Mitsuoka07,Ntshangase07} 
for systems relevant to the superheavy nuclei. 

In this article, 
in order to discuss the reaction dynamics for superheavy elements, 
we present the coupled-channels analyses for the 
quasi-elastic barrier distribution for the 
$^{48}$Ca+$^{248}$Cm system \cite{Tanaka18}. We consider the deformation 
of the target nucleus, $^{248}$Cm, as well as a neutron transfer channel. 
We also discuss the connection of the barrier distribution to  
the measured evaporation 
residue cross sections. 

\section{Quasi-elastic barrier distribution}

\begin{figure}[bt]
\begin{center}
\includegraphics[width=1.0\linewidth]{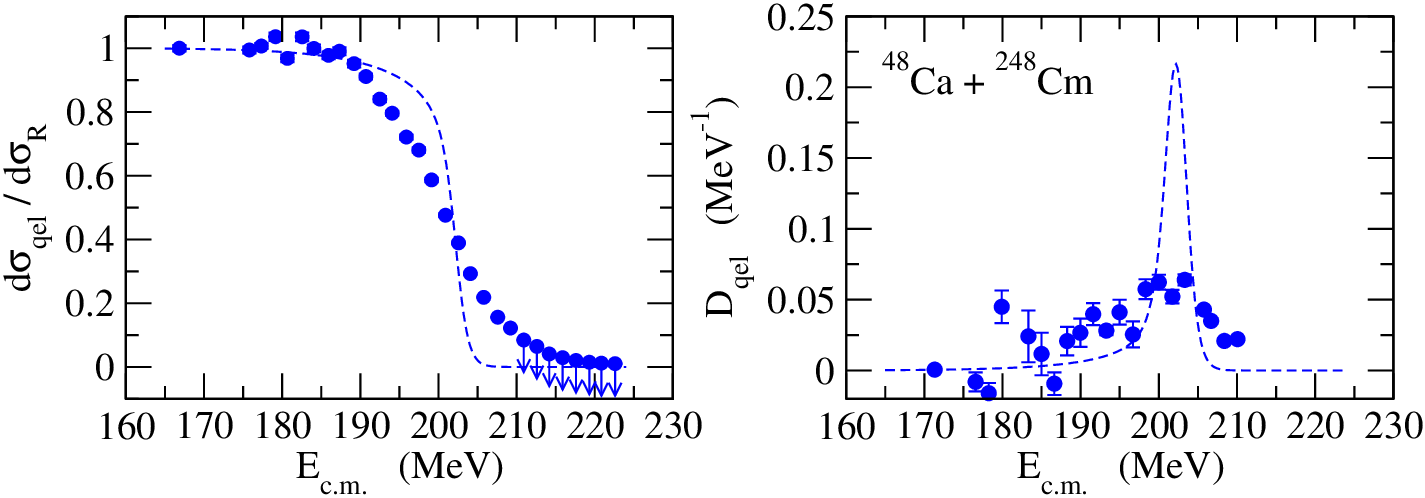}
\end{center}
\caption{The quasi-elastic cross sections (the left panel) and the 
corresponding barrier distribution (the right panel) 
for the $^{48}$Ca+$^{248}$Cm system. 
The experimental data are shown by the filled 
circles, where the arrows indicate the upper limit of cross sections. 
For comparison, the result of a potential model calculation is 
also shown by the dashed lines. The experimental data are 
taken from Ref. \cite{Tanaka18}. 
}
\end{figure}

Quasi-elastic scattering is defined as a sum of elastic, inelastic, 
transfer, and breakup processes. That is, it is an inclusive process, 
being complementary to fusion. The quasi-elastic barrier 
distribution, $D_{\rm qel}$, is defined as \cite{TLD95,HR04}, 
\begin{equation}
D_{\rm qel}(E)=-\frac{d}{dE}\left(\frac{\sigma_{\rm qel}(E,\pi)}{\sigma_R(E,\pi)}
\right), 
\end{equation}
where $\sigma_{\rm qel}(E,\pi)$ is a quasi-elastic cross section 
at the scattering angle 
$\pi$ and $\sigma_R(E,\pi)$ is the Rutherford cross section. 
In previous measurements, quasi-elastic scattering was measured by 
detecting projectile-like particles at the scattering angle $\theta$ 
(in the center of mass frame) and the cross sections were mapped on those at 
$\theta=\pi$ using the effective energy defined by \cite{TLD95,HR04}, 
\begin{equation}
E_{\rm eff}=2E\,\frac{\sin(\theta /2)}{1+\sin(\theta /2)}. 
\end{equation}
The mapping has been shown to work well as long as the scattering angle 
$\theta$ is close to $\pi$ \cite{HR04}. 
In the new measurement with GARIS, on the other hand, the recoiled target-like 
particles were measured at $\pi$ and the mapping of the quasi-elastic cross sections 
was not 
required \cite{Tanaka18}. 
This was not possible in the previous measurements, as it would have required 
to put a detector 
along the beam line. 
The measured quasi-elastic cross sections and 
the corresponding barrier distribution are shown in Fig. 1. 
The filled circles with arrows indicate the upper limit of cross sections, 
for which the deep-inelastic component may contaminate to some extent. 
For comparison, 
the figure also shows the result of a potential model. 
To this end, we use a Woods-Saxon potential with the depth parameter of 
$V_0=-105$ MeV, the range parameter of $r_0=1.18$ fm, and the surface 
diffuseness parameter of $a=0.6$ fm. The imaginary part of the potential 
is set to be well localized inside the barrier. 
It is clearly seen that the experimental barrier distribution is much more 
structured as compared to the result of the potential model, 
indicating the importance of channel coupling effects. 

\section{Coupled-channels calculations for the 
$^{48}$Ca+$^{248}$Cm system}

\begin{figure}[bt]
\begin{center}
\includegraphics[width=1.0\linewidth]{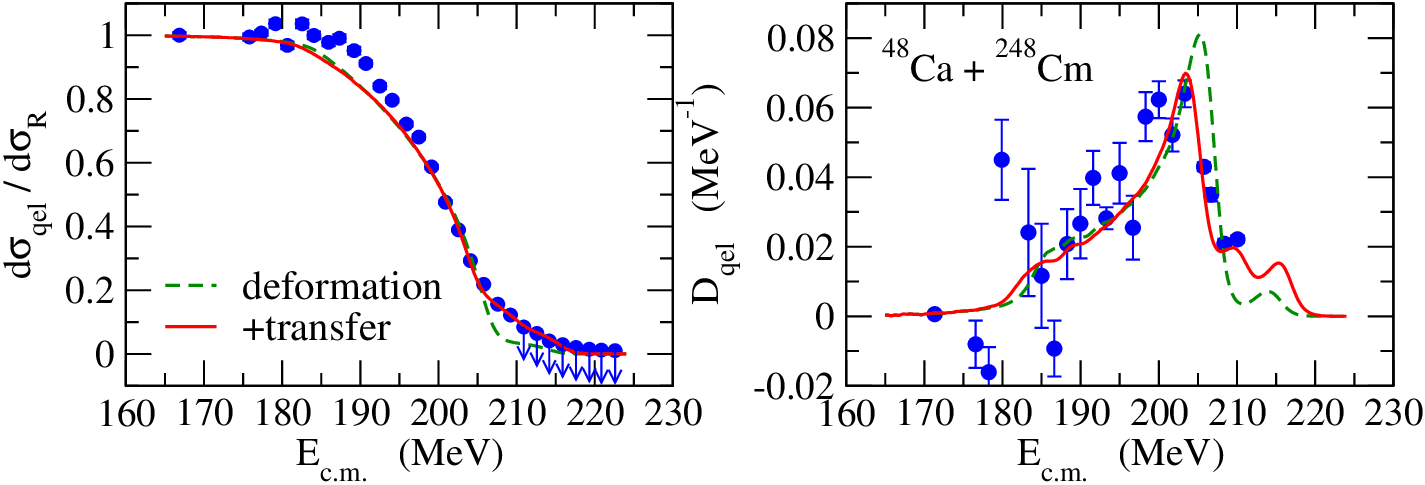}
\end{center}
\caption{
The quasi-elastic cross sections (the left panel) and the 
barrier distribution (the right panel) 
for the $^{48}$Ca+$^{248}$Cm system. 
The dashed lines show the result of coupled-channels calculations which 
take into account the deformation of the target nucleus, $^{248}$Cm, as 
well as the octupole phonon excitation in the projectile nucleus, $^{48}$Ca. 
The solid lines are obtained by including in addition the neutron transfer 
channel. 
The experimental data are 
taken from Ref. \cite{Tanaka18}. 
}
\end{figure}

Let us then perform coupled-channels calculations and discuss 
the role of channel coupling effects, in particular the role of 
deformation of the target nucleus, $^{248}$Cm. 
For a well deformed nucleus, for which the excitation energies of the 
ground state rotational band are small, cross sections for 
quasi-elastic scattering can be computed as \cite{HR04}, 
\begin{equation}
\frac{d\sigma_{\rm qel}}{d\Omega}=
\sum_I
\frac{d\sigma_I}{d\Omega}=
\int^1_0d(\cos\theta)
\left(\frac{d\sigma_{\rm el}}{d\Omega}\right)_\theta, 
\end{equation}
where $\sigma_I$ is the cross section to populate the state with spin $I$ 
in the ground state rotational band (including the ground state with $I=0$). 
$(d\sigma_{\rm el}/d\Omega)_\theta$ is the cross section for 
elastic scattering for a fixed orientation angle $\theta$ for the target 
nucleus. This is calculated with a deformed potential
for each $\theta$, $V(r,\theta)$, where $r$ 
is the relative distance between the two colliding nuclei. 
Notice that in this approximation the coupled-channels equations are 
completely decoupled and the cross sections are given as a weighted sum of 
cross sections for a single-channel system labeled by 
$\theta$ (for which there are only elastic 
scattering and absorption). 

The dashed lines in Fig. 2 show the result of a coupled-channels calculation 
so obtained. To this end, we use the deformation parameters of 
$\beta_2=0.297$ and $\beta_4=0.04$ together with the radius parameter 
of $R_T=1.2\times 248^{1/3}$ fm \cite{Hagino18}. 
We also take into account the octupole vibrational excitation in 
$^{48}$Ca at 4.51 MeV with the (dynamical) 
deformation parameter of $\beta_3=0.175$, although the excitation energy is 
large and 
the excitation simply renormalizes the internucleus 
potential \cite{HT12,THAB94}. 
One can see that a large part of the structure in the barrier distribution 
is well reproduced by this calculation. Especially, the asymmetric shape of the 
barrier distribution is well accounted for. 

A further improvement can be achieved by taking into account the 
neutron transfer channel. The solid lines in Fig. 2 show the results with 
the 1 neutron pick-up channel, whose ground-state-to-ground-state $Q$ value 
is $Q_{\rm gg}=-1.06$ MeV. 
We adjust the coupling strength for the transfer channel in order to 
reproduce the experimental quasi-elastic cross sections. 
This calculation slightly improves the cross sections around $E=210$ MeV, and 
the main peak in the barrier distribution is somewhat altered. 
However, the modification is minor, and one can conclude that 
the main effect still comes from 
the deformation of the target nucleus. 

\section{Connection to evaporation residue formations} 

\begin{figure}[bt]
\begin{center}
\includegraphics[width=0.5\linewidth]{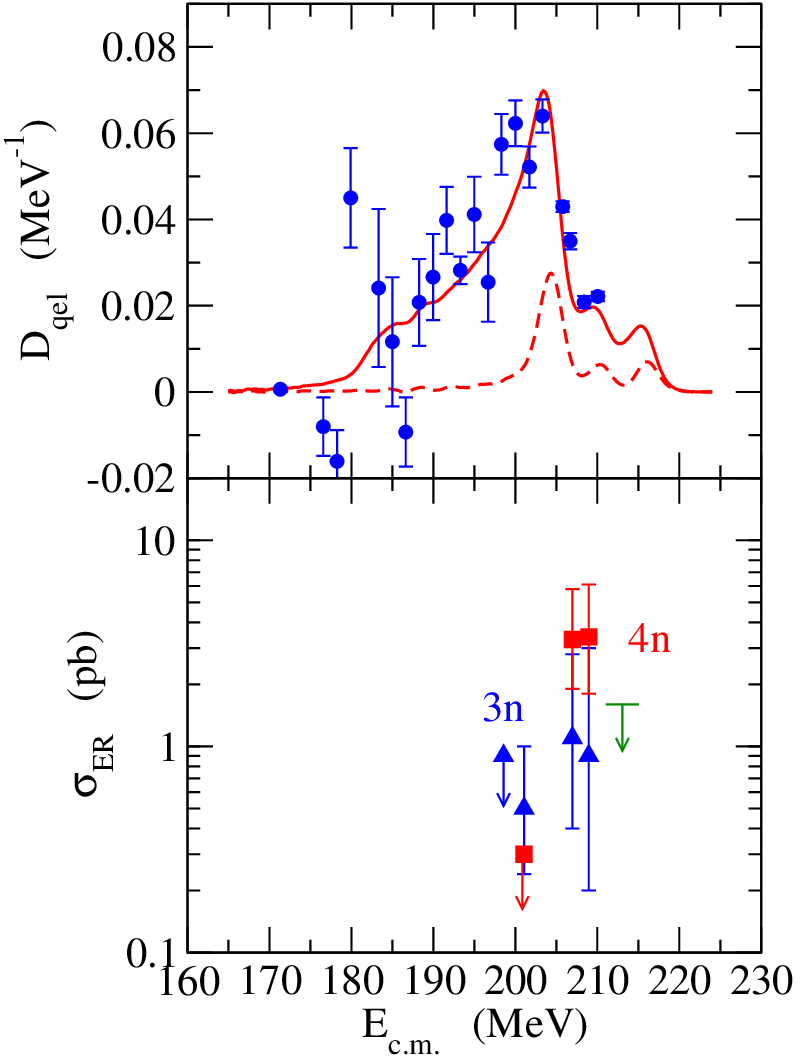}
\end{center}
\caption{The experimental quasi-elastic barrier distribution (the upper 
panel) and the evaporation residue cross sections (the lower panel)  
for the $^{48}$Ca+$^{248}$Cm system. 
For the barrier distribution, the figure also shows the results of the 
coupled-channels calculation (the solid line), for which the contribution 
of the side collision ($\theta=\pi/2$) is denoted by the dashed line. 
The experimental data are 
taken from Refs. \cite{Oganessian04,Hofmann12,Tanaka18}. 
}
\end{figure}

The barrier distribution discussed in the previous sections provides 
information on the Coulomb barrier in the entrance channel, that is, 
before the colliding nuclei reach the touching 
configuration. 
For medium-heavy systems, 
a compound nucleus is 
formed almost automatically once the touching configuration 
is achieved \cite{HT12}. 
In contrast, in the superheavy region, there is a 
huge probability for the touching configuration to 
re-separate without forming a compound nucleus, that is, quasi-fission, 
due to a strong Coulomb repulsion between the two nuclei. 
Furthermore, even if a compound nucleus is formed with a small 
probability, 
it decays most likely by fission. 
Because quasi-fission 
characteristics significantly overlap with fission of the compound nucleus, 
a detection of fission events itself does not guarantee a formation of the 
compound nucleus. Therefore, a formation of 
superheavy elements is usually identified by detecting 
evaporation residues. 
A question then arises: what is the connection between the barrier 
distribution in the entrance channel 
and evaporation residue cross sections? 

Figure 3 answers this question. This figure compares the quasi-elastic 
barrier distribution with the measured evaporation residue cross sections 
for the $^{48}$Ca+$^{248}$Cm system. 
For the barrier distribution, the figure also shows 
the result of the coupled-channels calculation (the solid line). The 
contribution of the side collision ($\theta=\pi/2$) is denoted by the dashed 
curve. 
The figure clearly indicates that 
the maximum of the evaporation residue cross sections originates from 
the side collision. 
This is a clear confirmation of the notion of compactness proposed by 
Hinde {\it et al.} \cite{Hinde95}, who argued that the side collision leads to 
a compact touching configuration, for which the effective barrier height 
for the diffusion process is low, enhancing the formation 
probability of a compound nucleus. 

\begin{figure}[bt]
\begin{center}
\includegraphics[width=0.5\linewidth]{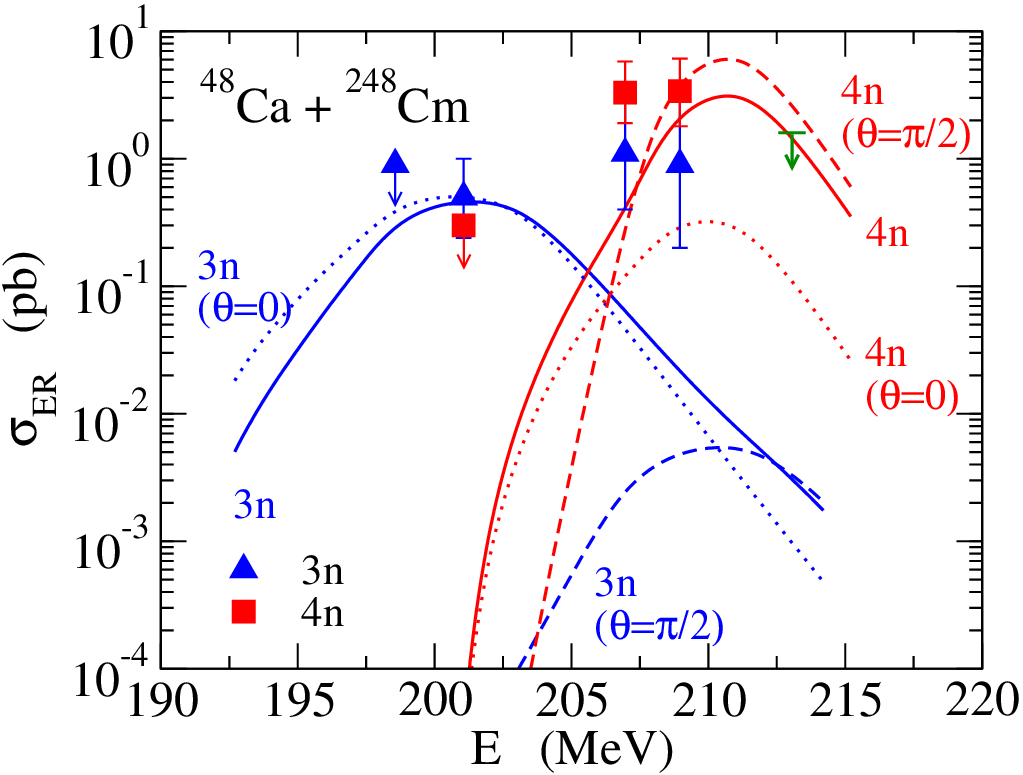}
\end{center}
\caption{The evaporation residue cross sections 
for the $^{48}$Ca+$^{248}$Cm system 
obtained with the 
extended fusion-by-diffusion model.  
The solid lines show the orientation averaged cross sections, while the 
dotted and the dashed lines denote the contributions of the tip ($\theta=0$) 
and the side ($\theta=\pi/2$) collisions, respectively. 
The experimental data are 
taken from Refs. \cite{Oganessian04,Hofmann12}. 
}
\end{figure}

This notion has further been confirmed theoretically 
using an extended version of the fusion-by-diffusion model \cite{Hagino18}. 
The fusion-by-diffusion model is a simple one-dimensional model 
for the diffusion process proposed 
by Swiatecki {\it et al.} \cite{fbd03,fbd05,fbd11}. 
In this model, the potential for the diffusion process is parameterized 
as a parabolic function of $s$, that is the surface separation between 
the two spheres. Assuming the overdamped limit, the diffusion 
probability, $P_{\rm CN}$, is then computed as \cite{Abe00}, 
\begin{equation}
P_{\rm CN}=\frac{1}{2}\left[1-{\rm erf}\left(\sqrt{\frac{\Delta V}{T}}\right)
\right],
\end{equation}
where ${\rm erf}(x)$ is the error function, $T$ is the temperature of the 
system, and $\Delta V=V_{\rm fiss}(s_{\rm sd})-V_{\rm fiss}(s_{\rm inj})$ 
is the difference between the 
potential energy at the saddle configuration, $s_{\rm sd}$, and that at 
the injection point, $s_{\rm inj}$. 
In the fusion-by-diffusion model, the fission potential, $V_{\rm fiss}(s)$, 
as well as the saddle configuration, $s_{\rm sd}$, are 
globally parameterized according to Refs. \cite{fbd03,fbd05,fbd11}, and the 
injection point, $s_{\rm inj}$, is treated as an adjustable parameter. 
In Ref. \cite{Hagino18}, this model has been extended by introducing the 
angle dependence to the injection point as, 
\begin{equation}
s_{\rm inj}(\theta)
=s_{\rm inj}^{(0)}+R_T\sum_\lambda\beta_{\lambda T}Y_{\lambda 0}(\theta). 
\end{equation}
The solid lines in Fig. 4 show the evaporation residue cross sections for the 
$^{48}$Ca+$^{248}$Cm system obtained with this model \cite{Hagino18}. 
The contributions of the side ($\theta=\pi/2$) and the tip ($\theta=0$) 
collisions are denoted by the dashed and the dotted lines, respectively. 
At energies around 211 MeV, that is, the energies slightly above the 
barrier height for the side collision (see Fig. 3), the side collision 
gives the main contribution. At these energies, the contribution of the 
tip collision is much smaller, since the injection point is large, and thus 
the diffusion probability is small. On the other hand, the side collision 
is largely suppressed at lower energies due to the small 
capture probability originated from a high capture barrier. The 
contribution of the tip collision then becomes dominant, for which the capture 
probability is not suppressed, even though the diffusion probability is small. 
In this way, the maximum of the evaporation residue cross sections are obtained 
at energies slightly above the barrier height for the side collision, at which 
the energy is high enough so that the capture probability is not suppressed 
and at the same time one can take advantage of large diffusion probabilities 
for the side collision. 

\section{Summary}

The fusion barrier distribution has provided useful information 
on the reaction dynamics for heavy-ion sub-barrier 
fusion reactions for many systems. 
This continues to be the case also for systems relevant to 
superheavy elements. 
In this article, we have discussed the quasi-elastic barrier distribution 
for the 
$^{48}$Ca+$^{248}$Cm system, that is, the system to synthesize the element 
116 (Lv) with hot fusion reaction. The coupled-channels analyses for the 
recently measured data have clearly indicated that the side collision 
plays an important role in forming evaporation residues and that the maximum 
of the evaporation residue cross sections appears at energies slightly above 
the height of the Coulomb barrier for the side collision. 
This has made a clear confirmation of the notion of compactness for the 
side collision, which has also been confirmed theoretically using the 
extended fusion-by-diffusion model. 

Of course, there still remain many challenges in nuclear reaction studies for 
superheavy elements, such as a clarification of shape evolution towards 
a compound nucleus with a deformed target, a role of quantum friction, 
and to understand the reaction dynamics of neutron-rich 
nuclei \cite{Hagino18-2}. 
Apparently much more theoretical and experimental works will be required in 
order to gain a deeper insight into reaction dynamics for superheavy elements. 

\ack

The authors thank K. Morita for useful discussions. 
K.H. also thanks V. Guimar\~aes for his hospitality during his stay in 
Sao Paulo.

\end{document}